\def\ee{{e^{i \theta_k}}}
\def\emf{{e^{-i\theta_k}}}

\def\bra{\langle}
\def\ket{\rangle}

\newcommand{\ourhat}[1]{\widehat{#1}}

\def\cpa{\left\{
\sum_{k}
\frac{1}{i+d_k}
\right\}^{-1}}
\def\cpad{\left\{
\sum_{k}
\frac{1}{-i+d_k}
\right\}^{-1}}

\def\cpaII{\left\{
\sum_{k}
\frac{1}{1-i d_k}
\right\}^{-1}}
\def\cpadII{\left\{
\sum_{k}
\frac{1}{1+i d_k}
\right\}^{-1}}
\def\RRn{{R_n\nu}}
\def\RRm{{R_m\mu}}

\documentstyle[prl,aps,epsfig]{revtex}
\begin{document}
\twocolumn[
\hsize\textwidth\columnwidth\hsize\csname@twocolumnfalse\endcsname

\title{\Large Cellular Dynamical Mean Field Approach to Strongly Correlated 
        Systems}
\author{Gabriel Kotliar, Sergej Y.~Savrasov and Gunnar P\'alsson}
\address{Serin Physics Laboratory, Rutgers University, Piscataway, New
Jersey 08855-0849, USA}
\date{September 22, 2000}

\maketitle
\begin{abstract}
  We propose a cellular version of dynamical-mean field theory which gives a
  natural generalization of its original single-site construction and is
  formulated in different sets of variables. We show how non-orthogonality of
  the tight-binding basis sets enters the problem and prove that the resulting
  equations lead to {\it manifestly causal} self energies.
\end{abstract}
\pacs{71.10.-w,~71.27.+a,~75.20.Hr}
]

Dynamical mean field theory (DMFT) has been very successful in describing many
aspects of strongly correlated electron systems \cite{review}, and presently
much effort is put into implementing it for realistic calculations of
materials properties of solids.  By construction, this method describes
correctly local correlations but misses altogether the effect of short-range
order.  This can be understood by noticing that in the limit of large lattice
coordination ($z \rightarrow \infty$) \cite{vollhardt} where the single-site
DMFT is exact, the interactions which induce short-range correlations, such as
magnetic superexchange, are scaled as $ 1\over z $ and in the absence of
frozen correlations with $z$ neighbors they disappear from the problem.  The
effects of magnetic correlations on single-particle properties are captured by
single site DMFT in magnetically {\it ordered} phases \cite{chitra,fleck}.
However many applications require generalizations of the DMFT to capture
short-range correlations, in the absence of broken symmetries. This is an
active area of research and several methods have already been put forward
\cite{review,ingersent,jarrell,cox3,katsnelson} for this purpose.

The intuitive idea behind cluster methods is to treat certain
local degrees of freedom (cluster degrees of freedom) exactly
while replacing the remaining degrees of freedom by a bath of non
interacting electrons which hybridize with the the cluster
degrees of freedom  so as to restore the translation invariance of
the original problem. The simplest example of this idea is the
Bethe Peierls cluster.
The basic difficulty of a
cluster technique is determining a suitable self consistency
condition  without generating unphysical solutions which violate
causality.  For a discussion of this longstanding problem in the
context of disordered systems see \cite{leath}. Ingersent and
Schiller\cite{ingersent} and independently Georges and
Kotliar\cite{review}, introduced a truncation of the skeleton
expansion in real space, which can be represented by solving
coupled impurity models of different sizes. This method is not
manifestly causal, but recent work \cite{cox3}, suggests that the
problems with causality encountered in the earlier treatments are
the result of inaccurate approximations in the solutions of the
impurity models. Jarrell \cite{jarrell} and collaborators have
suggested an alternative cluster scheme, the dynamic cluster
approximation (DCA) which is a cluster scheme in momentum space,
whereby the cluster considered, if regarded in real space has
periodic boundary conditions.  This body of work  \cite{jarrell} and  its
extensions 
\cite{katsnelson} 
established the computational feasibility and the existence of an a
priori causal cluster schemes.

In this paper we pursue   generalizations of the single site DMFT
inspired by analogies with electronic structure methods. This cellular
DMFT (CDMFT), remains close in spirit to the DMFT ideas described in
the introduction, where the clusters have free (and not periodic)
boundary conditions. We prove two central points:

\noindent 
I) The DMFT construction 
\cite{review} can be carried out  in a large
class of basis sets. This observation
frees us from the need to introduce  sharp boundaries 
in   real space.  This approach is inspired by ideas
from  electronic structure, in which one achieves a cellular
description by means of orbitals which can have  a variable spatial
extension.

\noindent 
II) This  CDMFT construction is manifestly
causal, i.e.~the self energies that result from the solution of the
cluster equations obey $\mbox{Im}\Sigma(k,\omega) \leq 0$, eliminating
a priori one of the main difficulties encountered earlier in devising
practical cluster schemes.

\begin{figure}[h]
\centering \epsfig{file=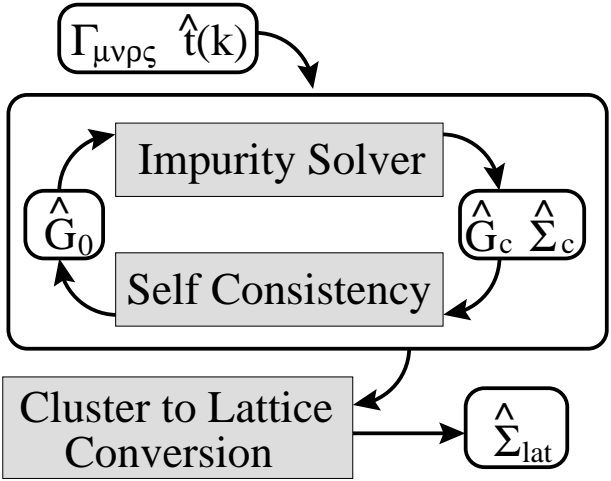,angle=-90,width=3.2in}
\caption{ Schematic separation of the elements of a cluster
dynamical mean field algorithm.  } \label{cluster.eps}
\end{figure}

\def\ff{f_{i\sigma}}
\def\ffd{f^{\dagger}_{i \sigma}}
It is useful to to separate the three essential elements of a
cluster scheme (see Fig.~\ref{cluster.eps}):

\noindent
a) The definition of the cluster degrees of freedom, which are
represented by impurity degrees of freedom in a bath described by a
Weiss field matrix function $\ourhat{G}_0$. The solution of the
cluster embedded in a medium results in a cluster Green's function
matrix and a cluster self energy matrix.

\noindent b) The expression of the Weiss field in terms of the
Green's function or the self energy of the cluster, i.e.~the
self-consistency condition of the cluster scheme.

\noindent c) The connection between the cluster self energy and
the self energy of the lattice problem.  The impurity solver
estimates the local correlations of the cluster, while the
lattice self energy is projected out using additional
information, i.e.  the periodicity of the original lattice.

Our construction applies to very general models for which a lattice
formulation naturally appears. It can be thought of as an extension of
the band structure formalism that takes into account the electron
electron interactions. The lattice Hamiltonian, $H[\ff,\ffd]$, (one
example could be the well known Hubbard Hamiltonian) is expressed in
terms of creation and annihilation operators $\ff$ and $\ffd$ where
$i$ runs over the sites of a d dimensional infinite lattice
$i=(i_1,\ldots,i_d)$, the index $\sigma$ denotes an internal degree of
freedom such as a spin index or a spin-orbital or band index if we
consider an orbitally degenerate solid.

\noindent
{\it a) Selection of cluster variables:} The first step in a mean
field approach to a physical problem, is a selection of a finite set
of relevant variables.  This is done by splitting the original lattice
into clusters of size $\prod_{j=1}^d$L$_j$ arranged on a superlattice
with translation vectors $R_n$.  On this superlattice we choose wave
functions $|R_n\alpha\ket$ partially localized around $R_n$ with
$\alpha = 1,\ldots,N$ denoting an internal {\it cluster} index.  The
relation between the new wave functions, $|R_n\alpha\ket$, and the old
ones, $|i\sigma\ket$, is encoded in a transformation matrix,
$S_{R_n\alpha,i\sigma}$, such that $|R_n\alpha\ket =
\sum_{i\sigma}|i\sigma\ket S^{-1}_{i\sigma,R_n\alpha}$.  Due to
the translation symmetry of the lattices we have
$S_{R_n\alpha,i\sigma} = S_{\alpha\sigma}(r(i)-R_n)$ where $r(i)$ is
the position of site $i$. The creation and annihilation operators of
the new basis are related to the operators of the old basis by
$c_{R_n\alpha} = \sum_{i\sigma}S_{R_n\alpha,i\sigma}f_{i\sigma}$ and
the operators that contain the "local" information that we want to
focus our attention on are $c_{\alpha} \equiv c_{(R_n=0)\alpha}$,
i.e.~the operators of the cluster at the origin.  We will refer to
these operators as the cluster operators.  Note that we do not require
that the wave function basis is orthogonal, and the nonorthogonality
is summarized in an overlap matrix $O^{mn}_{\mu\nu} =
O_{\mu\nu}(R_m-R_n) \equiv \bra R_m\mu|R_n\nu\ket$.
The next step is to express the Hamiltonian in terms of the complete
set of operators $c_{R_m\mu}$.  In terms of the new set of variables
it has the form
\begin{eqnarray}
H &=& -\sum_{R_m\mu R_n\nu}t_{\mu\nu}(R_m - R_n)
c_{R_m\mu}^{+}c_{R_n\nu} \nonumber \\
&+&\sum_{R_1\mu R_2\nu R_3\rho R_4\varsigma}
U_{\mu\nu\rho\varsigma}(\{ R_i \}) c_{R_1\mu
}^{+} c_{R_2\nu}^{+} c_{R_4\varsigma} c_{R_3\rho}.
\label{hubbard}
\end{eqnarray}
We stress again that the generality of the method. Equation
(\ref{hubbard}) has the form one would obtain by writing the full
Hamiltonian of electrons in a solid in some tight binding
non-orthogonal basis.  The Hamiltonian is now split into three parts,
$H = H_c + H_{cb} +H_b$ where $H_c$ involves only the cluster
operators, $H_b$ contains $c_{R_n\mu}$ with
$R_n\neq 0$ only and plays the role of a "bath", and finally $H_{cb}$
contains both $c_{R_n\mu}$ with $R\neq 0$ and the cluster operators
$c_\mu$.  Physically $H_{cb}$ couples the cluster with its
environment. A similar separation can be carried out at the level of
the action, in the coherent state functional integral formulation of
this problem, where the partition function and the correlation
functions are represented as averages over Grassman variables,
\begin{equation}
Z = \int \prod_{R_n\alpha} Dc^{+}_{R_n\alpha} Dc_{R_n\alpha} e^{-S}
\end{equation}
where the action is given by
\begin{eqnarray}
S &=&\int^{\beta}_{o}\!\! d\tau
\left(
\sum_{R_m\mu R_n\nu}\!c^{+}_{\RRm}O^{mn}_{\mu\nu}\partial_{\tau}c_{\RRn}
- H[c^{+}_{\RRm}, c_{\RRn}]
\right)
\nonumber \\
&\equiv& S_{c}+S_{cb}+S_{b}.
\label{Shubbard}
\end{eqnarray}
The effective action for the cluster degrees of freedom is obtained
conceptually by integrating out all the variables $c_{R_n\mu}$ with
$R_n \neq 0$ in a path integral to obtain an effective action for the
cluster variables $c_{\mu}$, i.e.
\begin{equation}
{{1}\over{Z_{eff}}} e^{-S_{eff} [c^{+}_{\mu} c_{\mu}]}
\equiv {{1}\over{Z}} \int \prod_{R_m\neq 0,\mu} Dc^{+}_{R_m\mu}
Dc_{R_m\mu} e^{ - S }.
\label{mf1}
\label{defseff}
\end{equation}
Note that the exact knowledge of $S_{eff}$ allows us to calculate {\it
  all the local} correlation functions involving cluster operators.
  As described in
\cite{review}, this cavity construction if carried out exactly
would generate terms of arbitrary high order in the cluster
variables.  Our approximation neglects the renormalization of  the
quartic  and higher order terms. Since the action $S_{cb}$
contains only boundary terms, the effects of these operators will
decrease as the size of the cluster increases.  Within these
assumptions, the effective action is parameterized by
$G_{0,\mu\nu}(\tau-\tau') $, the Weiss function of the cluster
and has the form
\begin{eqnarray}
\label{Seff}
S_{eff} &=& - \int^{\beta}_{0}\!\!\!\! d\tau d\tau'
\sum_{\mu\nu} c^{+}_{\mu}(\tau)G_{0,\mu\nu}^{-1}(\tau-\tau')
{c_{\nu}}(\tau')
\\
&&\hspace{-8ex} +
\int^{\beta}_{0}\!\!\!\! d\tau_1d\tau_2d\tau_3d\tau_4
\Gamma_{\mu\nu\rho\varsigma}
c^{+}_{\mu}(\tau_1)c^{+}_{\nu}(\tau_2)
c_{\varsigma}(\tau_4)c_{\rho}(\tau_3)
\nonumber
\end{eqnarray}
where $\Gamma_{\mu\nu\rho\varsigma} = U_{\mu\nu\rho\varsigma}(\{0\})$.
Using the effective action (\ref{Seff}) one can calculate the Green's
functions of the cluster $G_{c,\mu\nu}(\tau-\tau')[\ourhat{G}_0]
\equiv -\langle T_{\tau}c_{\mu}(\tau)c^{+}_{\nu}(\tau')\rangle
[\ourhat{G}_0]$
and the cluster self energies
\begin{equation}
\ourhat{\Sigma}_c \equiv \ourhat{G}_0^{-1}
 - \ourhat{G}_c^{-1}.
\end{equation}

\noindent 
{\it b) Self-consistency condition:} The cluster
algorithm is fully defined once a self-consistency condition
which indicates how $\ourhat{G}_0$ should be obtained from
$\ourhat{\Sigma}_c $ and $\ourhat{G}_c$ is defined.  In the
approach that we propose here the self consistent equations
become matrix equations expressing the Weiss field in terms of
the cluster self energy matrix $\ourhat{\Sigma}_c $.
\begin{equation}
\ourhat{G}_{0}^{-1} = \left(
\sum_{k}\frac{1}{(i\omega+\mu)\ourhat{O}(k)
-\ourhat{t}(k)-\ourhat{\Sigma}_c}\right)^{-1}+\ourhat{\Sigma}_c
\label{scc}
\end{equation}
where $\ourhat{O}(k)$ is the Fourier transform of the overlap matrix,
$\ourhat{t}(\mathbf{k})$, is the Fourier transform of the kinetic
energy term of the Hamiltonian in Eq.~(\ref{hubbard}) and {\bf k} is
now a vector in the Brillouin zone (reduced by the size of the
cluster, L$_j$, in each direction). Equations (\ref{Seff}) and
(\ref{scc}), can be derived by scaling the hopping between the
supercells as the square root of the coordination raised to power of
the Manhattan distance between the supercells and generalizing the
cavity construction of the DMFT \cite{review} from scalar to matrix
self energies. If the cluster is defined in real space and the self
energy matrices could be taken to be cyclic in the cluster indices so
that the matrix equations could be diagonalized in a cluster momentum
basis, Eq.~(\ref{scc}) would reduce to the DCA
equation\cite{jarrell}. However, in the DMFT construction, the
clusters have free and not periodic boundary conditions, and we treat
a more complicated problem requiring additional matrix inversions.
 
\noindent 
{\it c) Connection to the self energy of the lattice:} The
self-consistent solution, $\ourhat{G}_{c}$ and $\ourhat{\Sigma}_{c}$,
of the cluster problem can be related to the correlation functions of
the original lattice problem through the transformation matrix
$S_{R_m\alpha,i\sigma}$ by the equation
\begin{equation}
\Sigma_{lat,\sigma\sigma'}(k,\omega) = \sum_{\mu\nu}
\tilde{S}^{\dagger}_{\sigma,\mu}(k) \Sigma_{c,\mu\nu}(\omega)
\tilde{S}_{\nu,\sigma'}(k) \label{eight}
\end{equation}
where $\tilde{S}$ is the Fourier transform of the matrix S with
respect to the original lattice indices i. Notice that
$\Sigma_{lat,\sigma\sigma'}$ is diagonal in momentum and will also be
diagonal in the variable $\sigma$ if this variable is conserved.

\noindent {\it d) Connection to impurity models:} As in single
site DMFT it is very convenient to view the cluster action as
arising from a Hamiltonian,
\begin{eqnarray} \label{e2}
H_{imp} &=& \sum_{\rho\varsigma} \ourhat{E}_{\rho\varsigma}c^+_{\rho}
 c_{\varsigma} + \sum_{\mu\nu\rho\varsigma}
 \Gamma_{\mu\nu\rho\varsigma}c^+_{\mu}c^+_{\nu}c_{\rho}c_{\varsigma}
\nonumber \\
&+&
\sum_{k\lambda} \epsilon_{k\lambda}a^+_{k\lambda} a_{k\lambda}
+ \sum_{k\lambda,\mu} \left( V_{k\lambda,\mu}a^+_{k\lambda}c_{\mu} + h.c.
\right).
\end{eqnarray}
Here $\epsilon_{k\lambda}$ is the dispersion of the auxiliary band
and $V_{k\lambda,\mu}$ are the hybridization matrix elements
describing the effect of the medium on the impurity.  When the band
degrees of freedom are integrated out the effect of the medium is
parameterized by a hybridization function,
\begin{equation}
\Delta_{\mu\nu}(i\omega_n)[\epsilon_{k\lambda},V_{k\lambda}] =
\sum_{k\lambda}\frac{V^*_{k\lambda,\mu}V_{k\lambda,\nu}}{i\omega_n
  -\epsilon_{k\lambda}}.
\end{equation}
The hybridization function is  related to the Weiss field
function by expanding Eq.~\ref{scc} in high frequencies:
\begin{equation}
\label{eq:Weiss}
\ourhat{G}_0^{-1}(i\omega_n)
 =i\omega_n\bar{O}-\ourhat{E}-
\ourhat{\Delta}(i\omega_n)
\end{equation}
with $\bar{O}= \left[\sum_k\ourhat{O}_k^{-1}\right]^{-1}$ indicating
that the impurity model has been written in a non-orthogonal local
basis with an overlap matrix $\bar{O}$.

Let us now consider some examples of this approach.

\noindent 
{\it a) Single-site DMFT:} The simplest example is the single-site
dynamical mean field theory which is exact in the limit of infinite
dimensions.  In this case the cluster is just a single site denoted by
0, and the cluster operators are the creation and annihilation
operators of that site $ c^{+}_{o\sigma}$, $ c_{o\sigma}$. The cluster
Hamiltonian is diagonal in the spin variables and reduces to the
effective action of the Anderson impurity model. The second step is a
scalar equation $ \ourhat{G}_0^{-1} =
\ourhat{G}_c^{-1}[G_0]+\ourhat{\Sigma}_c[G_0]$.  Finally the third
step identifies the self energy of the cluster with the lattice
self energy.

\noindent
{\it b) Free cluster:} The next example is a free cluster scheme for
the one band Hubbard model. The method divides the lattice into
supercells, and views each supercell as a complex "site" to which one
can apply ordinary DMFT.  Here $R_{n}$ is the supercell position and
$\alpha$ labels the different sites within the unit cell, and the
spin.  Introducing a spin label $\sigma$ and a supercell notation
where an atom is denoted by the supercell, $R_n$, and the position
inside the supercell, $l$; $\alpha=(\sigma',l)$ and
$S_{R_n\alpha,i\sigma} =
\delta_{\sigma,\sigma'} \delta_{R_n+l,r_i}$ is diagonal in spin
and position. In this case the overlap matrix is the identity.
This  real space  cluster method 
has been investigated using quantum Monte Carlo
methods (QMC) by Katsnelson and Lichtenstein \cite{private}.

\noindent
{\it c) Multiorbital DMFT in a non-orthogonal basis:}  Another
important special case of our general construction is the
implementation of single-site DMFT in a non-orthogonal basis. In this
case the supercell is a single site, but the wave functions defining
the cluster operators are chosen so that they are very localized in
real space.

In fact, an implementation of this method, in conjunction with a
generalization of the interpolative perturbation theory, as an
impurity solver, has resulted in new advances in the theory of
Plutonium \cite{savrasov} Here the flexibility in the choice of basis
is crucial for the success of the DMFT program.  DMFT neglects from
the start interactions which are not onsite. A high degree of
localization requires a non-orthogonal basis and the formalism
introduced in this letter.

\noindent
{\it d) Other bases:} Finally we point out that the most attractive
feature of this method is that it would allow its formulation in terms
of wave functions which are partially localized in real and momentum
space such as wavelet functions.  This flexibility is most appealing
for treating problems such as the Mott transition where both the
particle-like and the wave-like aspect of the electron need to be
taken into account requiring a simultaneous consideration of real and
momentum space.

We now prove that the CDMFT approach gives manifestly causal Green's
functions. For this we assume, that we start the DMFT iteration with a guess
for the bath function $\ourhat{\Delta}$ which is causal. The self energy which
is generated in the process of solving the "impurity model" is also causal.
Furthermore, any sensible approximation techniques to compute the self energy
of the cluster respects causality, so our proof is valid not just for exact
solutions of the CDMFT scheme but also for approximate solutions as long as
the impurity solvers used in the solution of the cluster impurity problem
preserve causality. The next step is to show that if a causal self energy is
introduced in the self-consistency condition, (\ref{scc}), the resulting bath
function $\ourhat{\Delta}$ is causal. Since both $\ourhat{\Sigma}_c$ and
$\ourhat{\Delta}$ are matrices, the causality condition needs to be formulated
precisely.  For a Fermionic matrix function, $A(\omega)$, to be causal means
that it is analytic in the upper half of the complex frequency plane and thus
has a spectral representation with spectral density $\frac{-1}{2\pi
  i}\{A(\omega)-A^{\dagger}(\omega)\}$, and the spectral density matrix is
positive definite.  It is easy to see that the DMFT equations lead to the
correct analytic properties and the following proof establishes the positivity
of the bath spectral density.  Writing $\Sigma_R = \epsilon-i\gamma$ with
$\epsilon$, $\gamma$ hermitian and $\gamma$ positive definite, we get:
%
\begin{eqnarray}
(\ourhat{\Delta}_R^{\dagger}-\ourhat{\Delta}_R) &=& -2i\gamma
\\
&& \hspace{-13ex}
+ \sqrt{\gamma}
\left(\cpa\hspace{-0.3cm}-\cpad\right)\sqrt{\gamma}.
\nonumber
\end{eqnarray}
Positivity is reduced to proving that the following matrix is negative
\cite{footnote},
i.e.
\begin{equation}
2- \cpaII\hspace{-0.3cm} - \cpadII \leq 0.
\label{positive}
\end{equation}
Here $\omega\ourhat{O}(k)-\ourhat{t}(k)-\epsilon \equiv
\sqrt{\gamma} d_k \sqrt{\gamma}$.  Performing a change of variables
$d_k =i {{ {\ee}+ {\emf}} \over {{\ee} - {\emf}}}$ with
${\theta_k}$ a hermitian matrix, Eq.~(\ref{positive}) reduces to
proving that
\begin{equation}
1 \leq   \min_x {{  \bra x | (1-z)^{-1} + (1-z^{\dagger})^{-1} | x \ket} \over {\bra x |x \ket}}
\label{zeq}
\end{equation}
where $z \equiv \langle\!\langle e^{-2 i{\theta_k}}\rangle\!\rangle$ ,
with $\langle\!\langle\rangle\!\rangle$ denoting an average over the
Brillouin zone.  By performing the substitution $| x \ket =
(1-z)|y\ket$ Eq.~(\ref{zeq}) reduces to
\begin{equation}
1 \leq \min_{||y|| = 1}\frac{ 2 +  \bra y | z + z^{\dagger} | y
\ket} {1 + \bra y| z^{\dagger} z | y\ket + \bra y | z +
z^{\dagger}|y \ket} \label{sixteen}
\end{equation}
which clearly holds due to the fact that $z $ is an average of unitary
matrices, and has the property $ ||z^{\dagger} z || \leq 1$.  From
Eq.~(\ref{sixteen}) and Eq.~(\ref{eight}) it follows that the imaginary part
of the retarded self energy is always less or equal to zero, completing the
proof of causality. This proof generalizes  Ref.~\cite{jarrell}
from scalar to matrix CPA equations.

In conclusion DMFT has produced a wealth of information in problems
where the physics is local and cluster methods promise to be equally
fruitful in more complex problems where correlations between more
sites and orbitals need to be taken into account.  All the techniques
which have been used for the solution of the single site DMFT are
applicable to this cluster extension.  The most powerful methods of
solution have been renormalization group related techniques such as
the projective self-consistent method
\cite{fisher}  or the numerical Wilson renormalization group
techniques \cite{bulla}. These methods carry out  a division of
the impurity and the bath into a low-energy and a high-energy
part, and perform an elimination of the high-energy degrees of
freedom in both.  Extensions of these methods to an abstract
cluster is a first step in constructing  {\it non-perturbative}
renormalization group method for correlated fermion systems and
deserves further investigations.


\begin{thebibliography}{99}
\bibitem{footnote} The matrix inequality  $ B \leq 1$
is defined  by the condition $\bra x|B|x \ket \leq \bra x |  x \ket  $
for all vectors x.
\end{thebibliography}
\end{document}